# STAR CLUSTER KINEMATICS WITH AAOMEGA

László L. Kiss (Univ. of Sydney), Zoltán Balog (Univ. of Arizona), Gyula M. Szabó (Univ. of Szeged, Hungary), Quentin A. Parker (Macquarie Uni. & AAO), David J. Frew (Perth Observatory)

**Introduction**

The high-resolution setup of the AAOmega spectrograph makes the instrument a unique stellar radial velocity machine, with which measuring Doppler shifts to $\pm 1.3$ km s$^{-1}$ for 16 magnitude stars within an hour of net integration has become a reality. The 1700D grating with its spectral resolution of $\lambda/\Delta\lambda = 10000$ in the near-infrared calcium triplet (CaT) range between 8400–8800 Å is particularly well suited for late-type stars, whose spectral energy distribution peaks exactly in this range and whose spectra are dominated by the strong CaT lines. The large field of view and the instrument's capabilities form an excellent combination for kinematic studies of star clusters.

The two types of stellar aggregates, open and globular clusters, are good representations of the two ends of stellar evolution and, as hosts to thousands of stars of the same age and chemical composition, they have played a key role in understanding stellar structure and evolution. An open cluster is a group of stars that were formed in the same giant molecular cloud and which are still bound together by a relatively weak gravitational field. Their ultimate fate is dispersal in the Milky Way field, illustrated by the fact that even the oldest known open clusters (e.g. M67, NGC 188) are generally only a couple of Gyr old. Globular clusters, on the other hand, are the remnants of the first galactic building blocks with ages from 10 to 13 Gyr, which means their old populations are the best known local counterparts of the high-redshift Universe.

Traditionally, the relationship between the colours and magnitudes of stars in a cluster has been used to derive fundamental parameters such as age, distance, reddening and metallicity of the given cluster. These can, in turn, reveal important details of stellar physics (Gallart et al. 2005). The spectroscopic approach, i.e. taking individual spectra of hundreds or thousands of stars and then deriving global parameters of the clusters from the spectral analysis of the member stars, has been much less utilised before the advent of efficient multi-object spectrographs. Enormous efforts have been undertaken in a few cases (e.g. Meylan & Mayor 1986) but star cluster kinematics was very far from routine in observational astronomy. The situation has completely changed in the last few years, when a number of new instruments came online (Hectospec/Hectoechelle at MMT, FLAMES at VLT, DEIMOS at Keck, AAOmega at AAT, etc.), delivering thousands of spectra at a speed and sensitivity never seen before.

Radial velocities tell a different story to the colour-magnitude diagram: velocity dispersion is linked to the total mass of the cluster, hence indicating the presence or absence of invisible matter; the dispersion as a function of radius is a tell-tale indicator of the underlying mass profile, whereas systemic rotation can be revealed through an analysis of angular distribution of the velocities. Coupled with proper motion measurements, velocities can also be used to derive a kinematic distance (assuming energy equipartition for the member stars). A completely new avenue opens up with the full spectral analysis, when atmospheric parameters such as effective temperature, surface gravity and metallicity, are determined for each star. In that case, evolutionary models can be fitted directly to the physical parameters rather than the colours and magnitudes, which are sensitive to the interstellar reddening.

In semester 2008A, Balog et al. were granted four nights of AAOmega time to observe the relatively young double open cluster NGC 2451A and B. These have been studied with the Spitzer space telescope to identify stars with infrared excess caused by circumstellar debris disks, which are thought to host on-going planet formation. In this article we report on the first results of the project based on three nights of observations from early 2008A. Due to the main target's limited visibility, we have also obtained data for secondary objects, including the apparent association of the open cluster M46 and the planetary nebula NGC 2438, and the pair of globular clusters near Antares, M4 and NGC 6144. The amazing results which came out of only three nights of AAT time illustrate very nicely the potential of the instrument and, for example, how quickly one can resolve decades of contradiction in less than two hours of net observing time.

**Observations and data analysis**

We acquired AAOmega data on three nights in February 2008, in moderate Siding Spring sky conditions. In the blue arm we used the 2500V grating, providing $\lambda/\Delta\lambda = 8000$ spectra between 4800 Å and 5150 Å. In the red arm we used the 1700D grating that has been optimised for recording the CaT region.

In total, we acquired 11 field configurations centered on NGC 2451, 2 configurations on M46 and 3 configurations on M4. The target stars were selected from the 2MASS point source catalogue (Skrutskie et al. 2006) by matching the main features in the colour-magnitude diagram of stars in each cluster. In every configuration we limited the brightness range of stars to 3 mag in order to avoid cross-talk between the fibres due to scattered light.

The spectra were reduced using the standard 2dF data reduction pipeline. We performed continuum normalisation separately for the stellar spectra using the IRAF task *onedspec.continuum* and then removed the residuals left from the strongest skylines by linearly interpolating the surrounding continuum.

Atmospheric parameters and radial velocity were determined for each star with an iterative process, which combined finding the best-fit synthetic spectrum from the Munari et al. (2005) spectrum library, with $\chi^2$ fitting, and cross-correlating the best-fit model with the observed spectrum to calculate the radial velocity. This approach is very similar to that adopted by the Radial Velocity Experiment (RAVE) project (Steinmetz et al. 2006; Zwitter et al. 2008), and this analysis is based on the same synthetic library used by RAVE. Our experiences have shown that because of the wide range of temperatures (and hence spectral features), we needed three subsequent iterations to converge to a stable set of temperatures, surface gravities, metallicities and radial velocities. The latter are believed to be accurate within ±1–2 km s⁻¹ for the cooler stars and ±5 km s⁻¹ for the hotter stars in the sample (the boundary is roughly at 8000–9000 K). These values have been estimated from Gaussian fits of the cross-correlation profile using the IRAF task *rv.fxcor* and should only be considered as representative numbers.

The specific uses of the data were as follows:

- For NGC 2451, we wanted to identify genuine cluster member stars in the photometrically pre-selected sample, based on their full set of parameters (radial velocities, temperature, surface gravities, metallicities).

- For M46 and NGC 2438, we wanted to confirm or rule out physical association between the cluster and the nebula, most notably by comparing their radial velocities.

- For M4 and NGC 6144, we wanted to measure velocity dispersion profiles near to or

beyond the tidal radii and estimate central velocity dispersion.

In the following three sections we describe the science case for each cluster and present preliminary results. More details will appear in Balog et al. (in prep.), Kiss et al. (2008, submitted) and Kiss et al. (in prep.).

### NGC 2451A and B: two open clusters in the same line-of-sight

Debris disks provide evidence for the presence of planetary objects around young stars. They form when large bodies collide, generating fragments that participate in cascades of further collisions resulting in a significant quantity of dust grains. These dust grains are heated by the central star and then re-radiate at longer wavelengths. This reprocessed radiation is detectable with Spitzer through excess emission at mid- and far-infrared wavelengths. The dust grains are relatively short-lived ($10^6$–$10^7$ yr), hence they must be regenerated by further collisions. Therefore, their presence is the strongest evidence of the existence of large bodies (up to planet size) that collide and produce dusty debris.

Debris disks provide a great opportunity to study how planetary systems form and to follow their evolution through time. It is important to have a sample of debris disk systems with well determined age. The aims of the Spitzer program No. 58 "Evolution and Lifetime of Protoplanetary Disks" are to investigate the frequency and duration of the protoplanetary disk phase of evolution and to obtain constraints on the probabilities and timescales for the formation of major planetary bodies.

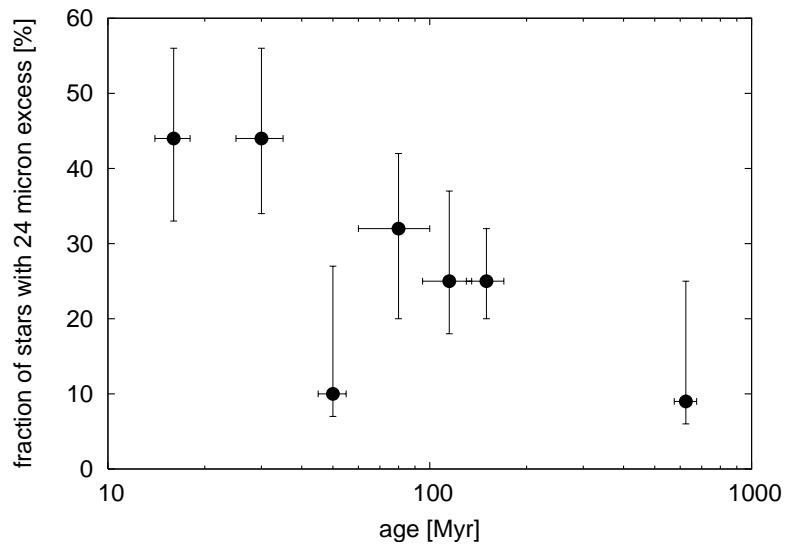

Fig.1: Frequency of stars B5-A9 with 24 micron excess fraction as a function of age (a simplified version of Fig. 5 in Siegler et al. (2007)).

Our motivation is to examine the timescale for and nature of the transition to the debris disk phase of disk evolution. To achieve these goals we surveyed a sample of young stellar clusters of varying age, richness and stellar content in the age range of 1–100 million years.

This survey has revealed a remarkable similarity between the general evolution of hundreds of planetary systems and our ideas about the early events in the Solar System. There is a decay in debris generation with a time scale of about 100 Myr. A few exceptional objects have been found that may signal major collisions such as the one between the proto-Earth and a large planetesimal that resulted in the formation of the Moon. Debris in the terrestrial planet zones (detected at 24 microns) persists for about 1 Gyr, parallel to the 700 Myr period that ended in the Late Heavy Bombardment in the Solar System. Therefore, we are sampling many planetary systems in a way that will let us test and expand our current theories for planet evolution and will also show us whether some of the salient events in the development of the Solar System are common or rare. However, there is an important gap in our time coverage between 30 and 100 Myr that needs to be filled to deliver on the full promise of the Spitzer data. NGC 2451 A and B are ideal for filling this gap (Fig. 1).

NGC 2451A and B are two young open clusters projected onto each other in the same line-of-sight. Several attempts have been made to separate the two clusters and to determine the physical parameters of each. Platais et al. (2001) analysed photometric and spectroscopic data and used proper motion, radial velocities to select members of NGC 2451A. They fitted theoretical isochrones to the cluster colour-magnitude diagram (CMD) to calculate distance, reddening and age, deriving d=188 pc, E(B-V)=0.01 mag and t=60 Myr. Hünsch et al. (2003) carried out an X-ray study of the two clusters. These authors identified 39 members of the A and 39 of the B cluster, using combined X-ray and optical data, determined distances of 206 pc and 370 pc for NGC 2451A and B, respectively, and found ages around 50–80 Myr for NGC 2451A and about 50 Myr for NGC 2451B. Subsequently, Hünsch et al. (2004) completed the X-ray study with high resolution spectroscopy and refined the membership of the two clusters. The most recent distance and age estimates were published by Kharchenko et al. (2005), who analysed the ASCC-2.5 catalog and provided homogeneous astrophysical parameters for 520 Galactic open clusters. They estimated distances of 188 pc and 430 pc and ages of 57.5 Myr and 75.9 Myr for NGC 2451 A and B, respectively.

The central 1 deg x 1 deg field of NGC2451 was imaged with Spitzer/IRAC (3.6, 4.5, 5.8 and 8.0 micron) and MIPS (24, 70, 160 micron). Supplementary observations in UBVRI bands were obtained with the 1.5m telescope at Las Campanas Observatory in Chile. Using the parameters of Kharchenko et al. (2005) and the V vs V-K colour-magnitude diagram, we attempted to separate the two clusters in order to investigate their stellar content and disk frequencies. However, our member selection method becomes uncertain around V-K~4, where the level of contaminating background red giants is very high. In the infrared, the background red giants can mimic the observational signatures of debris disks around low mass main sequence stars, so it is very important to separate them from the main sequence stars before we start identifying sources with debris disks in the clusters. That was the main goal of our AAOmega observations: to clean the sample of background red giants and thus refine membership determination, and also to separate the two clusters from each other.

In total, we determined radial velocity and atmospheric parameters for 2757 stars in the Spitzer field. The histogram of

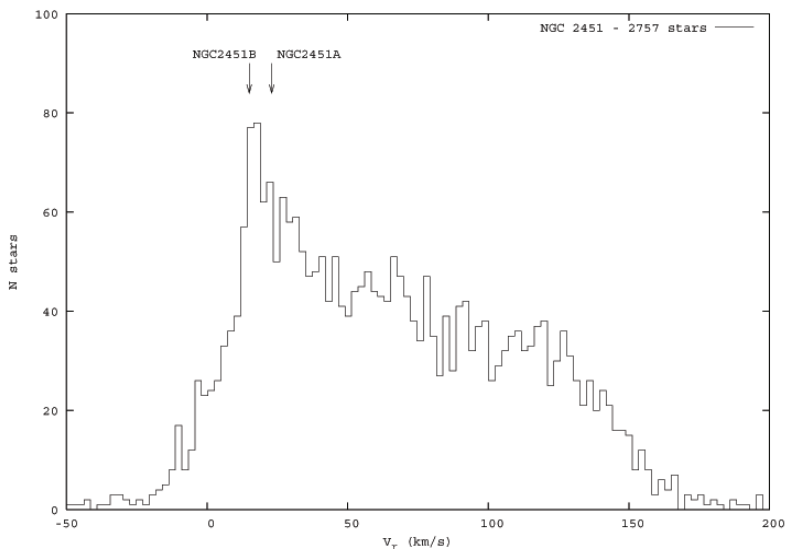

Fig. 2: The histogram of all radial velocities for NGC 2451. The two arrows show the published velocities of the two overlapping clusters (Hünsch et al. 2004). The overwhelming majority of the stars belong to the Milky Way field.

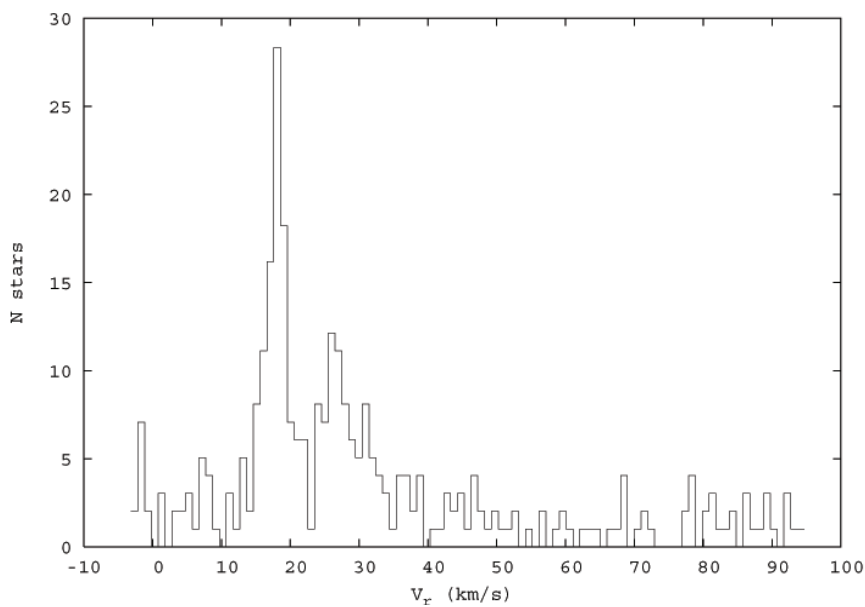

Fig. 3: The histogram of refined radial velocities for stars with distorted CaT line profiles. About 150 stars can be identified as members of either of the two clusters, with mean velocities of 18 km s⁻¹ and 26 km s⁻¹.

all radial velocities (Fig. 2) indicates a heavy contamination from field stars: the two clusters are barely noticeable at the published velocities (marked by two arrows). Another difficulty that arose was the very high incidence of CaT emission (defined by the excess flux of the calcium lines relative to the best-fit model spectra), especially in – as revealed later – late-type main sequence stars in the clusters. Both clusters are young, hence the still high rotation rates can lead to elevated chromospheric activity, for which the CaT lines are good indicators (Andretta et al. 2005). However, someone's signal is someone else's noise: the distorted CaT profiles introduced random velocity errors of up to 10–20 km s⁻¹ in the maximum of the cross-correlation profile. This was particularly apparent after improving the velocity determination for 312 stars with CaT line profile irregularities. After exclusion of the CaT lines from the cross-correlation, the velocity of a much larger fraction of the stars were very close to the clusters' mean. (Fig. 3)

At the time of writing this article, we have confirmed the existence of two clusters in the same line-of-sight based on the most extensive kinematic database obtained for NGC 2451. A large fraction of cool main sequence member stars has been identified with chromospheric activity, which alone will be the basis for further studies in an unanticipated direction. Currently we are working on the analysis of debris disk candidates, whose results will be presented in Balog et al. (in prep.).

**M46 and the planetary nebula NGC 2438**

Any physical associations discovered between planetary nebulae (PNe), the short-lived but spectacular late evolutionary stage of small and intermediate mass stars (between 1–8 $M_{sun}$), and star clusters would be a valuable discovery providing a means of establishing accurate astrophysical parameters for the nebulae through fixing distances and progenitor ages from cluster isochrones. Accurate distances are particularly useful, as from them one can infer PNe physical properties such as the absolute magnitude of the central stars, accurate physical dimensions and fluxes. Also, they would provide excellent calibrators for the surface brightness-radius relation (Frew & Parker 2006, Frew 2008). Whereas PNe have been found in 4 globular clusters of the Milky Way (M15, M22, Pal 6 and NGC~6441; Jacoby et al. 1997), none has been reported in the literature as an unambiguous member of a much younger open cluster (OC). The interest in the latter case is not only due to being able to determine independent distances to individual nebulae, but also because in a young open cluster the progenitor of a now-visible PN will be a reasonably constrained higher mass star than those in globular clusters. This fact offers the opportunity to calibrate the initial-to-final mass relation of stars on a broad range of masses, usually done by modelling white dwarf populations in open clusters (e.g. Dobbie et al. 2006).

Recently, Majaess et al. (2007) and Bonatto et al. (2008) have performed detailed investigations of possible physical associations between PNe and OCs. Majaess et al. considered the cluster membership for 13 PNe that are located in close proximity to open clusters lying in their lines-of-sight and listed another 16 PNe/open cluster coincidences, which might contain physically associated pairs. However, they noted that we have yet to establish a single association between a PN and an open cluster based on a correlation between their full set of physical parameters, including the three key parameters of radial velocity, reddening, and distance that need to be in good agreement if an association is to be viable. Bonatto et al. (2008) used near-infrared colour-magnitude diagrams and stellar radial density profiles to study PN/open cluster association for four pairs. They concluded that the best, but still only probable, cases are those of NGC 2438/M46 and PK 167-01/New Cluster 1.

NGC 2438 is a well-known annular PN located about 8 arcminutes from the core of the bright open cluster M46 (=NGC2437, Back page). Despite its brightness, the cluster was relatively unstudied until recently (e.g. Cuffey 1941; Stetson 1981). Recently published cluster parameters are relatively well-determined, e.g. E(B-V)=0.10–0.15, D =1.5–1.7 kpc and an age of 220–250 Myr (Sharma et al. 2006, Majaess et al. 2007, Bonatto et al. 2008). The estimated turnoff mass is about 3.5 $M_{sun}$. In addition to the possible association with NGC 2438, M46 is also thought to host the well-studied post-AGB candidate OH 231.8+4.2 (Jura & Morris 1985).

Early studies of the radial velocity of NGC 2438 and M46 (Cuffey 1941; O'Dell 1963) indicated a difference of about 30 km s[-1] between the PN and cluster stars, which suggested that the pair constitutes a spatial coincidence only. Three red giants in the cluster have systemic velocities (Mermilliod et al. 1989, 2007) identical to that of cluster dwarf members obtained by Cuffey (1941). However, Pauls & Kohoutek (1996) rekindled interest in the possibility of the PN/open cluster association when they found similar velocities for both, although based on a small number of stars. Both Majaess et al. (2007) and Bonatto et al. (2008) pointed out the importance of measuring sufficient stellar radial velocities for the cluster and the PN to establish if the proximity is real or only chance superposition. Prompted by this recent interest, we observed two configurations on M46 and three positions across the face of the PN (Fig. 4) to resolve the ambiguities and confusion in the literature.

In total, 105 min of integration with two configurations led to radial velocities of 586 stars in a 1 deg field of view centered on M46. Unlike NGC 2451, where chromospheric activity of late type stars led to difficulties in velocity determination, here it was the hot early type cluster members that caused a bit of extra work. We show sample spectrum fits in Fig. 5 to illustrate the difficulties one faces when analysing cool and hot stars together in the CaT region. Since the CaT lines almost exactly coincide with hydrogen lines in the Paschen series, we found that it was absolutely crucial to have the best-match template for cross-correlation. A slight template mismatch can easily lead to radial velocity shifts of several km s$^{-1}$ at this intermediate spectral resolution and hence one has to be very careful to optimise template selection. It is common practice in the optical range to use the same template across a range of spectral subtypes or even types. However, that does not work in the CaT region, where a full $\chi^2$ fit of the spectra is essential. It is also inevitable that as soon as the temperature reaches about 9000 K, the broad spectral features will lead to a degraded velocity precision simply because of the broadened cross-correlation profile. M46, as an intermediate-age open cluster, still hosts a large number of hotter main sequence stars and that implies the possibility of degraded velocity precision for a significant fraction of stars. But as we show below, we confirm the 30 km s$^{-1}$ velocity difference between the cluster and the PN, so that the temperature dependent velocity uncertainty does not play a role in relation to their physical association.

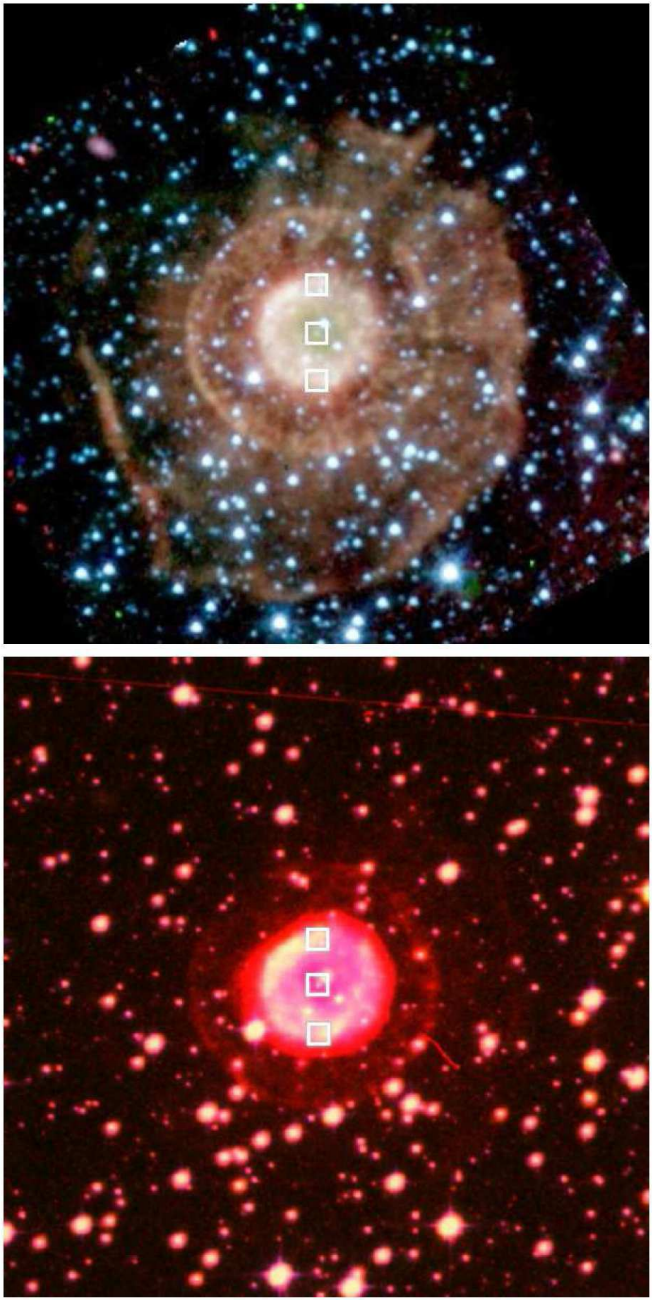

Fig. 4: Spitzer/IRAC (upper panel) and SuperCOSMOS Hα (lower panel) images of NGC 2438. The white squares show the three fibre positions we used to take spectra of the PN. The field of view is 6 arc minutes.

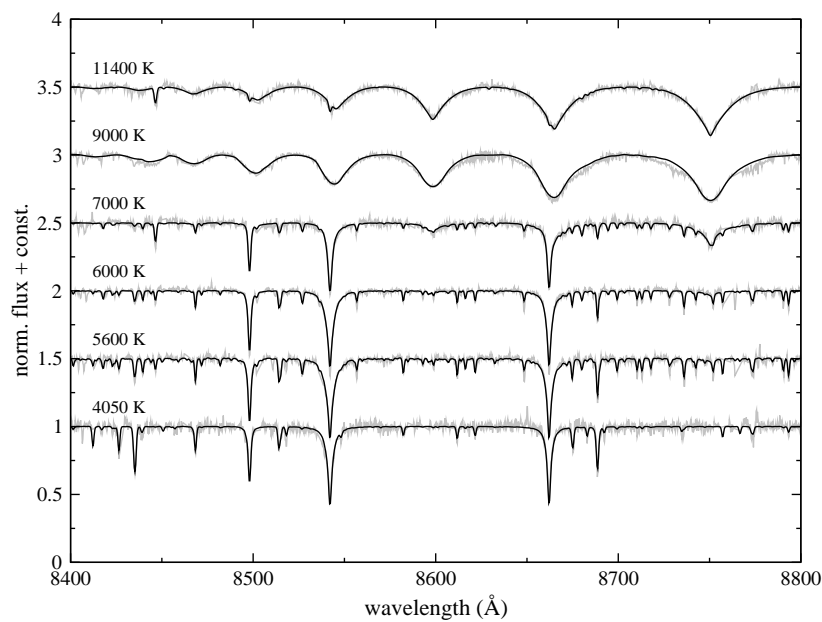

Fig. 5: Observed stellar spectra (light blue lines) and the best-fit synthetic data from the Munari et al. (2005) spectrum library (black lines).

The spectra taken in the three positions across NGC2438 are typical of a planetary nebula. The blue range shows the three characteristic nebular emission lines, the H$\beta$ and the [O$_{III}$] doublet at 4959 Å and 5007 Å, which are by far the strongest features in the optical spectrum. The central spectrum shows well-defined double-peaked [O$_{III}$] line profiles, which we used to determine the expansion velocity of the nebula, assuming a spherical shell. The two spectra on the edge have single-peaked emissions, centered exactly halfway between the two peaks of the central spectrum, supporting that assumption.

In the near-IR, the central position yielded a featureless flat continuum, while the two spectra from the shell contain an identical set of narrow emission lines. Using line identifications from the literature we identified these lines as the Paschen series of hydrogen (from P12 to P18), the [Cl$_{III}$] nebular line at 8579 Å, and the N$_{I}$ line at 8729 Å.

Individual PN radial velocities have been measured by fitting Gaussian functions to the line profiles. In the case of the double-peaked [O$_{III}$] doublet, we fitted a sum of two Gaussians. In each case we repeated the centroid measurement by choosing slightly different fit boundaries to estimate the uncertainty: the strong emission lines in the blue yielded the same velocities within 1–2 km s$^{-1}$ in several repeats. The mean velocity of the PN from our data is 78±2 km s$^{-1}$, while the [O$_{III}$] expansion velocity is 21.0±0.2 km s$^{-1}$, both in excellent agreement with numbers in the literature (e.g. Corradi et al. 2000).

Fig. 6 shows the histogram of the most extensive and accurate set of radial velocities for the open cluster M46 obtained to date. We confirm the early results that NGC 2438 has a relative velocity of about 30 km s$^{-1}$ with respect to the cluster (O'Dell 1963), hence they are unrelated despite being located approximately at the same ~1.5–1.7 kpc distance. In Fig. 6 we also put an arrow at the mean centre-of-mass velocity (48.5 km s⁻¹) of three red giant binaries measured by Mermilliod et al. (1989, 2007). The excellent agreement between the maximum of the histogram (49 km s⁻¹ for the highest value, 48 km s⁻¹ for the centroid of the fitted Gaussian) and the very accurate CORAVEL data confirms both the cluster membership of those systems and the quoted accuracy of our single-epoch velocity measurements.

At the time of writing, we have therefore ruled out a physical association between the open cluster M46 and the planetary nebula NGC 2438. We also noted the very broad velocity peak of the cluster in the histogram, for which the presence of a significant population of binary stars has been concluded. More details can be found in Kiss et al. (2008, submitted).

### Two flies in one hit: the globular clusters M4 and NGC 6144

We were also able to observe M4, within the field of which lies another globular cluster, NGC 6144. In semesters 2006B and 2007B we were granted 15 nights to observe globular clusters (first results published by Kiss et al. 2007 and Székely et al. 2007) and now we have added these two clusters to the six already observed (47 Tuc, M12, M30, M55, NGC 288, NGC 6752) .

Globular clusters are spherical aggregates of 10⁴–10⁶ very old stars bound in regions as small as a few tens of parsecs. They are believed to have undergone substantial dynamical evolution because of long ages compared to the relaxation time scale. This evolution is affected by several different processes including tidal interaction with the Galaxy and two-body relaxation, which are both responsible for the "evaporation" of stars (Meylan & Heggie 1997). A globular cluster that moves on a non-circular orbit around the Galactic centre experiences a time-dependent tidal force. When the duration of this perturbation is much shorter than the typical orbital periods of stars within the cluster, stars experience an abrupt tidal shock that can have dramatic effects on the evolution of the cluster (Gnedin et al. 1999). These shocks occur when the cluster passes through the galactic disk or the bulge; the disk shock compresses the cluster while the bulge shock stretches it. The effects of such a shock continue long after the shock itself: the energy given to the individual stars in the cluster accelerates both the general evolution and mass evaporation. The stars therefore escape continuously and, because of the non-spherical equipotential surfaces, the cluster is expected to have tidal tails. N-body simulations of globular clusters have shown that tidal tails may be used to trace the orbital paths of globular clusters and hence give direct information on the Galactic gravitational field and the underlying mass distribution.

Deep star-count surveys have revealed tidal tails in two low-concentration clusters (Palomar 5 and NGC 5466) stretching many degrees beyond the tidal radius (Odenkirchen et al. 2001; Grillmair & Johnson 2006), confirming the theoretical predictions. The AAOmega instrument on the AAT offers wonderful new opportunities to investigate mechanisms that affect velocity distributions in globular clusters and, in particular, near or outside the tidal radius. Theories to be tested include tidal heating of the evaporated stars by the external gravitational field (Drukier et al. 1998), the presence of a dark matter halo around the clusters (Carraro & Lia 2000), and a breakdown of the Newtonian dynamics in the weak-acceleration regime (Scarpa et al. 2007). The latter hypothesis is particularly interesting because modified Newtonian dynamics, valid for accelerations below $a_0 \sim 1.2 \times 10^{-8}$ cm s$^{-2}$, may offer an alternative to dark matter, with far-reaching implications for cosmology.

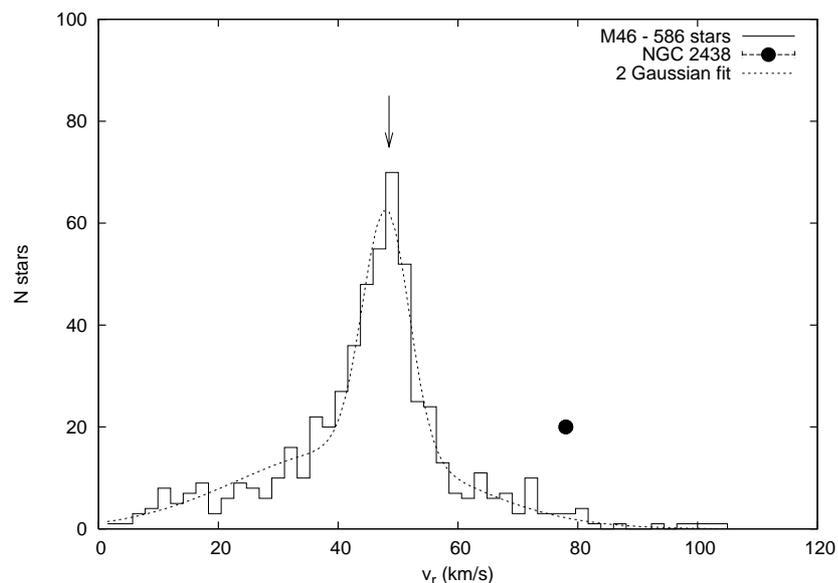

Fig. 6: The histogram of stellar radial velocities for M46. The arrow shows the mean center-of-mass velocity of three red giant binaries (Mermilliod et al. 1989, 2007), while the dotted line represents a fitted sum of two Gaussians. About half of the sample belongs to the cluster.

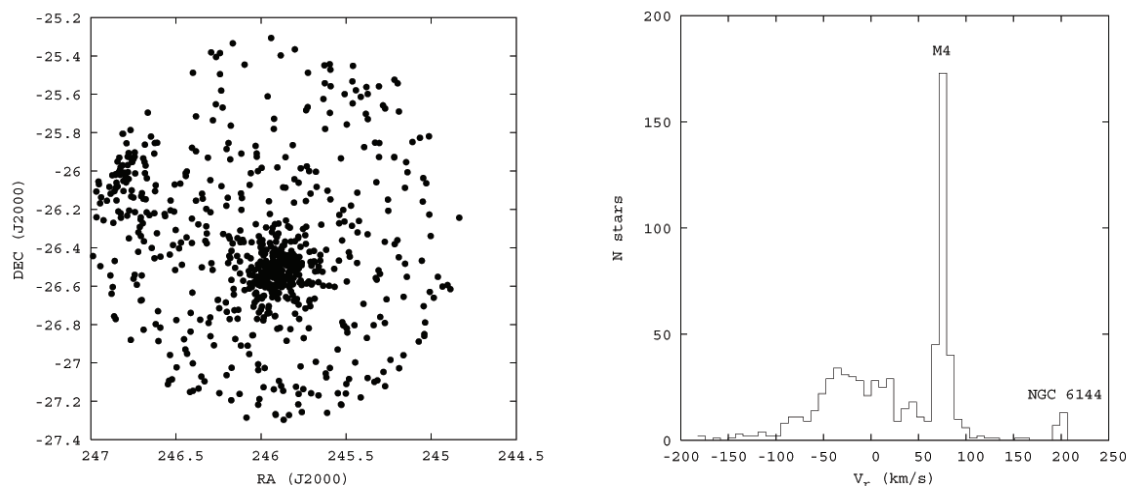

Fig. 8: Left: the celestial distribution of the M4 sample. Note the concentration on the left edge, which marks the location of NGC 6144. Right: the histogram of radial velocities with the two cluster peaks marked.

We therefore aimed at recording radial velocities for as many cluster member stars as possible, located everywhere from the centre to beyond the tidal radii. In the case of M4/NGC 6144, we have acquired three field configurations containing candidate red giant cluster members. In Fig. 7 we show the celestial distribution of the 719 stars observed (left panel) and the histogram of the measured radial velocities (right panel). The two peaks of the two clusters at 75 km s$^{-1}$ (M4) and 200 km s$^{-1}$ (NGC 6144) are very well defined and distinct from the galactic field.

A preliminary analysis shows that approximately 300 members can be identified in M4, whose radial velocities indicate a detectable systemic rotation with a full amplitude of about 1 km s$^{-1}$ and a surprisingly large velocity dispersion around the tidal radius. The detailed analysis is underway and the results will be published later this year (Kiss et al., in prep.).

# Are M46 and NGC2438 associated?
# AAOmega provides a definitive answer

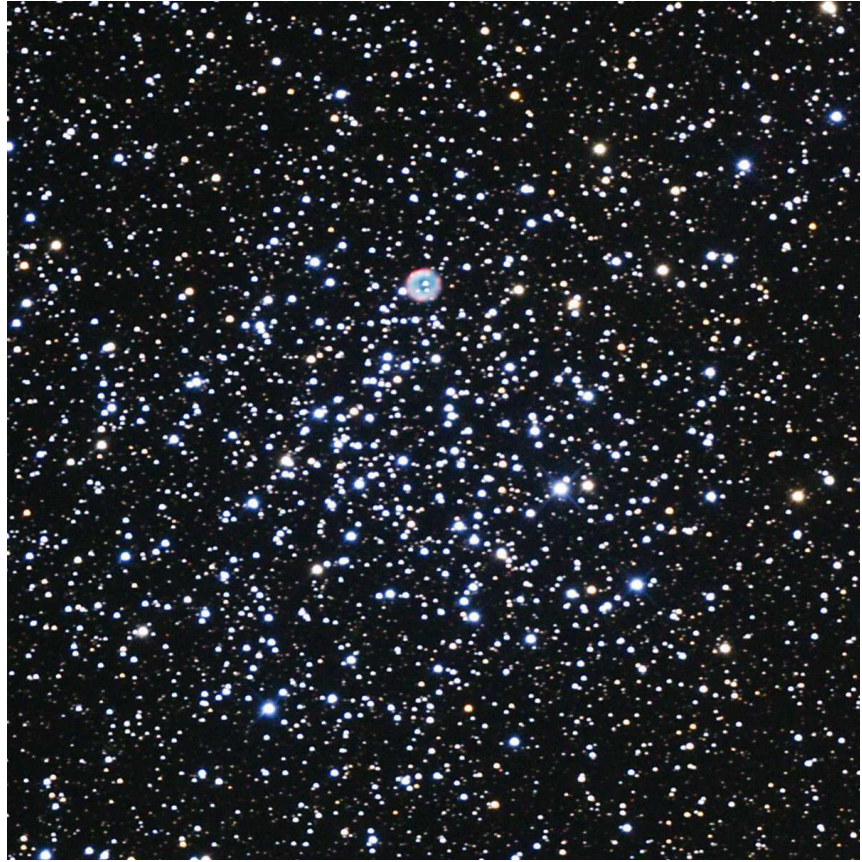

A colour image (courtesy of Steve Lee) of the 250 Myr-old cluster M46 and the planetary nebula NGC2438. Until now there has been uncertainty as to whether these two objects are physically associated. In this issue László Kiss and collaborators describe their recent work on radial velocities with AAOmega, which has provided a definitive answer to this question.

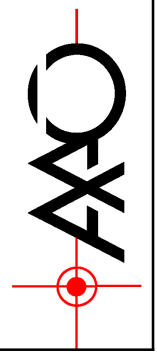